\documentstyle{article}
\input psfig

\def\MPL #1 #2 #3 {Mod.~Phys.~Lett.~{\bf#1},\  #2 (#3)}
\def\NPB #1 #2 #3 {Nucl.~Phys.~{\bf#1},\  #2 (#3)}
\def\PLB #1 #2 #3 {Phys.~Lett.~{\bf#1},\  #2 (#3)}
\def\PR #1 #2 #3 {Phys.~Rep.~{\bf#1},\ #2 (#3)}
\def\PRD #1 #2 #3 {Phys.~Rev.~{\bf#1},\  #2 (#3)}
\def\PRL #1 #2 #3 {Phys.~Rev.~Lett.~{\bf#1},\  #2 (#3)}
\def\RMP #1 #2 #3 {Rev.~Mod.~Phys.~{\bf#1},\  #2 (#3)}
\def\ZP #1 #2 #3 {Z.~Phys.~{\bf#1},\  #2 (#3)}
\def\IJMP #1 #2 #3 {Int.~J.~Mod.~Phys.~{\bf#1},\  #2 (#3)}
\def\hpm{H^{\pm}}
\def\hmm{\Delta^{--}}
\def\mhmm{m_{\hmm}}

\def\gamhmm{\Gamma_{\hmm}}

\def\wtil{\widetilde}

\def\shat{{\hat s}}
\def\rtshat{\sqrt{\shat}}

\def\sigrts{\sigma_{\!\!\rts}^{}}

\def\nsigrts{n_{\sigrts}}

\def\sighbar{\overline \sigma_{\h}}

\def\anti{\overline}

\def\zstar{Z^\star}
\def\wstar{W^\star}

\def\mupmum{\mu^+\mu^-}

\def\rts{\sqrt s}

\def\eg{{\it e.g.}}

\def\anti{\overline}
\def\wp{W^+}
\def\wm{W^-}
\def\mw{m_W}
\def\mz{m_Z}
\def\h{h}           % clever macro
\def\mh{m_{\h}}
\def\gamh{\Gamma_{\h}^{\rm tot}}
\def\hsm{h_{SM}}
\def\mhsm{m_{\hsm}}
\def\gamhsm{\Gamma_{\hsm}^{\rm tot}}
\def\tanb{\tan\beta}
\def\hl{h^0}
\def\mhl{m_{\hl}}

\def\ha{A^0}
\def\mha{m_{\ha}}

\def\hh{H^0}
\def\mhh{m_{\hh}}

\def\fbi{~{\rm fb}^{-1}}
\def\fb{~{\rm fb}}
\def\mev{~{\rm MeV}}
\def\gev{~{\rm GeV}}
\def\tev{~{\rm TeV}}
\def\mstop{m_{\wtil t}}
\def\mt{m_t}
\def\mb{m_b}

\def\mm{\mu^+\mu^-}
\def\ee{e^+e^-}

\def\tanb{\tan\beta}

\def\overlay#1#2{\ifmmode \setbox 0=\hbox {$#1$}\setbox 1=\hbox to\wd 0{\hss
$#2$\hss }\else \setbox 0=\hbox {#1}\setbox 1=\hbox to\wd 0{\hss
#2\hss }\fi #1\hskip -\wd 0\box 1}
\def\case#1/#2{{\textstyle{#1\over#2}}}
\def\9{\phantom 0}      %%% for lining up numbers in columns
\renewcommand\linebreak{\unskip\break} %% breaks line & still justifies
\newcommand{\alt}{\mathrel{\raisebox{-.6ex}{$\stackrel{\textstyle<}{\sim}$}}}
\newcommand{\agt}{\mathrel{\raisebox{-.6ex}{$\stackrel{\textstyle>}{\sim}$}}}
\def\lsim{\alt}
\def\gsim{\agt}
\catcode`@=11
\newcount\@tempcntc
\def\@citex[#1]#2{\if@filesw\immediate\write\@auxout{\string\citation{#2}}\fi
  \@tempcnta\z@\@tempcntb\m@ne\def\@citea{}\@cite{\@for\@citeb:=#2\do
    {\@ifundefined
       {b@\@citeb}{\@citeo\@tempcntb\m@ne\@citea\def\@citea{,}{\bf ?}\@warning
       {Citation `\@citeb' on page \thepage \space undefined}}%
    {\setbox\z@\hbox{\global\@tempcntc0\csname b@\@citeb\endcsname\relax}%
     \ifnum\@tempcntc=\z@ \@citeo\@tempcntb\m@ne
       \@citea\def\@citea{,}\hbox{\csname b@\@citeb\endcsname}%
     \else
      \advance\@tempcntb\@ne
      \ifnum\@tempcntb=\@tempcntc
      \else\advance\@tempcntb\m@ne\@citeo
      \@tempcnta\@tempcntc\@tempcntb\@tempcntc\fi\fi}}\@citeo}{#1}}
\def\@citeo{\ifnum\@tempcnta>\@tempcntb\else\@citea\def\@citea{,}%
  \ifnum\@tempcnta=\@tempcntb\the\@tempcnta\else
   {\advance\@tempcnta\@ne\ifnum\@tempcnta=\@tempcntb \else \def\@citea{--}\fi
    \advance\@tempcnta\m@ne\the\@tempcnta\@citea\the\@tempcntb}\fi\fi}
\catcode`@=12

\renewenvironment{thebibliography}[1]
 {\begin{list}{\arabic{enumi}.}
    {\usecounter{enumi} \setlength{\parsep}{0pt}
     \setlength{\itemsep}{3pt} \settowidth{\labelwidth}{#1.}
     \sloppy
    }}{\end{list}}

\begin{document}

\newlength{\captsize} \let\captsize=\small % use \let\normalsize=\captsize
%\newlength{\captwidth}                     % just before \caption{ ...

%Draft \today

\twocolumn[
\font\fortssbx=cmssbx10 scaled \magstep1
\hbox to \hsize{
%\special{psfile=uwlogo.ps hscale=5000 vscale=5000 hoffset=-12 voffset=-2}
%\hskip.25in\raise.05in
\hbox{\fortssbx University of Wisconsin - Madison}
\hfill$\vcenter{\hbox{\bf MADPH-96-939}
            \hbox{\bf UCD-96-12}
            \hbox{April 1996}}$}

\medskip

\begin{flushleft}
{\Large\bf
Particle Physics Opportunities at
{\protect\boldmath$\mu^+\mu^-$} Colliders}\footnotemark\\
\medskip
V.~Barger$^a$, M.S.~Berger$^b$, J.F.~Gunion$^c$,
T.~Han$^c$\\
\medskip
$^a$Physics Department, University of Wisconsin, Madison, WI 53706, USA\\
\smallskip
$^b$Physics Department, Indiana University, Bloomington, IN 47405, USA\\
\smallskip
$^c$Physics Department, University of California,  Davis, CA 95616, USA\\
\end{flushleft}
\vspace*{-.25in}
]

\footnotetext{Based on talks given at the {\it Symposium on Physics Potential and Development of $\mu^+\mu^-$ Colliders}, San Francisco, California, December 13--15, 1995.}
%\begin{abstract}  no special format or title for Abstract in SF report

We discuss the capabilities of future muon colliders to resolve important
particle physics questions. A collider with c.m.\ energy $\sqrt s = 100$ to
500~GeV offers the unique opportunity to produce Higgs bosons in the
$s$-channel and thereby measure the Higgs masses, total widths 
and several partial widths to high
precision. At this same machine, $t\anti t$ and
$\wp\wm$ threshold studies would yield superior precision
in the determination of $\mt$ and $\mw$.
A multi-TeV $\mm$ collider would open up the realm of physics above
the 1 TeV scale, allowing, for example, 
copious production of supersymmetric particles
up to the highest anticipated masses or a detailed study
of the strongly-interacting scenario of electroweak symmetry breaking.

%\end{abstract}

\let\Large=\normalsize %% so section titles are normal size
\let\large=\normalsize %% ditto for subsections

\section{\uppercase{Introduction}}
\indent

There is increasing interest recently in the possible construction of a $\mm$
collider\cite{mupmumi,saus,montauk,theseprocs}. 
The expectation is that a muon collider
with energy and integrated luminosity comparable to 
or superior to those attainable at $\ee$
colliders can be achieved\cite{palmer,neuffersaus,npsaus}.
An initial survey of the physics potential of muon colliders has been carried
out\cite{workgr}.
In this report we summarize some of the progress on the physics issues that has
been made in the
last year; a more comprehensive report is in preparation\cite{mupmumreport}.

One of the primary arguments for an $e^+e^-$ collider is the complementarity
with physics studies at the LHC. The physics potential of a muon collider is
comparable to that of an electron collider with the same energy and luminosity.
However, electron colliders are at a technologically more advanced stage and
will likely be built before muon colliders.  Hence a very relevant issue is
what can be done at a muon collider that cannot be done at an electron
collider.

The advantages of a muon collider can be summarized briefly as follows:

\begin{itemize}

\item The muon is significantly heavier than the electron, and therefore
couplings to Higgs bosons are enhanced making possible their study in the
$s$-channel production process.

\item The limitation on luminosity from beam-beam interactions 
that arises at an $\ee$ collider is not relevant for muon beam
energies below about 100 TeV; very small/flat beams
are unnecessary. Instead, large luminosity is achieved for $\sim 3\,\mu$m
size beams by
storing multiple bunches in the final storage ring and having a large
number of turns of storage per cycle. 
Radiative losses in the storage ring are small due to the
large muon mass. Thus,
extending the energy reach of these colliders well beyond the 1 TeV range
is possible.

\item The muon collider 
can be designed to have finer energy resolution than an $\ee$ machine.

\item At a muon collider, $\mu^+\mu^+$ and $\mu^-\mu^-$ collisions are 
likely to be as easily achieved as $\mu^+\mu^-$ collisions.

\end{itemize}

\textheight21cm

There are two slight drawbacks of a muon collider.  The first
is that substantial polarization of the beams can probably not
be achieved without sacrificing luminosity. The second
drawback is that the $\gamma\gamma$ and $\mu\gamma$ options
are probably not feasible. At future linear $\ee$ colliders, the possibility
exists to
backscatter laser photons off the electron and/or positron beams.
The resulting back-scattered
photons are highly collimated and could serve as a photon beam, thus converting
the $\ee$ collider to a $e\gamma $ or $\gamma \gamma $ collider.
The collisions from the back-scattered photons have center-of-mass energies
that range up almost to that of the parent $\ee$ collider.
Including this option at a $\mm$ collider is problematic from kinematic
considerations. The highest photon energy $\omega $ attainable from a
lepton with energy $E$ is
\begin{equation}
{\omega_{\rm max}\over E}={x\over {x+1}}\;,
\end{equation}
where
\begin{equation}
x={4E\omega_0\over m_\mu^2c^4}\;.
\end{equation}
For a muon collider $x\ll 1$ unless a laser photon energy $\omega _0$
of the order of keV is possible, which seems unlikely.

\begin{figure}[t]
\let\normalsize=\captsize   %%%% changes the font to "\small"
\begin{center}
\centerline{\psfig{file=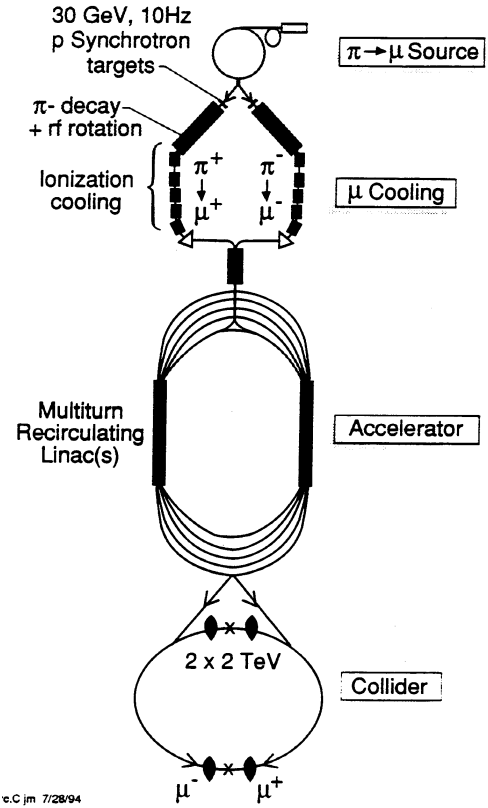,width=6.5cm}}
\begin{minipage}{7cm}       %%%% reduces width of caption to 7cm
\bigskip
\caption{\baselineskip=0pt
A possible design for a muon collider, from Ref.~\protect\cite{saus}.}
\label{collfig}
\end{minipage}
\end{center}
\end{figure}

A proposed schematic design for a muon collider is shown in
Fig.~\ref{collfig}. Protons
produce $\pi$'s in a fixed target which subsequently decay giving $\mu$'s.
The muons must be collected, cooled and subsequently accelerated to high
energies. Since the muon is so much heavier than the electron, synchrotron
radiation is much less so that circular storage rings are feasible even at TeV
energies.

The monochromaticity of the beams will prove critically important for some
of the physics that can be done at a $\mm$ collider. The energy profile of
the beam is expected to be roughly Gaussian in shape, and the rms
deviation $R$ is expected to naturally lie in the range
$R = 0.04$\% to 0.08\%\cite{jackson}. Additional cooling could further
sharpen the beam energy resolution to $R=0.01\%$.

%1

Two possible $\mm$ machines have been discussed as design targets and
are being actively studied \cite{saus,montauk,theseprocs}:

\begin{enumerate}
\item[(i)] A first muon collider (FMC) with
low c.~m.\ energy ($\rts$) between $100$ and $500\gev$ and
${\cal L}\sim2\times10^{33}\rm\,cm^{-2}\,s^{-1}$ delivering an annual
integrated integrated luminosity $L\sim 20\fbi$.

\item[(ii)] A next muon collider (NMC) with high $\rts \agt 4$ TeV and
${\cal L} \sim 10^{35}\rm\,cm^{-2}\,s^{-1}$ giving
$L\sim 1000\fbi$ yearly.
\end{enumerate}

\section{{\protect\boldmath$s$}-\uppercase{channel Higgs physics}}

The simplest Higgs sector is that of the
Standard Model (SM) with one Higgs boson. However, 
the naturalness and hierarchy problems that arise in the SM
and the failure of grand unification of
couplings in the SM suggest that
a single Higgs boson is probably not the
whole story of electroweak symmetry breaking. Therefore, it is crucially
important to understand and
delineate experimentally various alternative possibilities. 

Supersymmetry is an especially attractive candidate theory in that it solves
the naturalness and hierarchy problems (for a sufficiently low scale
of supersymmetry breaking) and in that scalar bosons, including
Higgs bosons, are on the same
footing as fermions as part of the particle spectrum.
The minimal supersymmetric model\break
 (MSSM) is the simplest SUSY extension
of the SM.  In the MSSM, every SM particle has a superpartner.
In addition, the minimal model contains exactly two Higgs doublets.
At least two Higgs doublet fields are required
in order that both up and down type quarks be given masses
without breaking supersymmetry (and also to avoid anomalies
in the theory).  Exactly two doublets allows unification of
the SU(3), SU(2) and U(1) coupling constants. 
(Extra Higgs singlet fields are allowed by unification, but are presumed
absent in the MSSM.) For two Higgs doublets and no Higgs singlets,
the Higgs spectrum comprises
5 physical Higgs bosons \begin{eqnarray}
&&\hl, \hh, \ha, H^+, H^-\;.
\end{eqnarray}
The quartic couplings in the MSSM Higgs potential are related to the
electroweak gauge couplings $g$ and $g'$ and the tree-level Higgs mass
formulas imply an upper bound on the mass of
the lightest Higgs boson, $m_h\leq M_Z$.
At one loop, the radiative correction to the mass of the 
lightest Higgs state depends on the top and stop masses
\begin{eqnarray}
&&\delta \mhl^2\simeq {{3g^2}\over {8\pi ^2\mw^2}}\mt^4
\ln \left ({{m_{\tilde{t}_1}m_{\tilde{t}_2}}\over {\mt^2}}\right )\;.
\end{eqnarray}
Two-loop corrections are also significant.
The resulting  ironclad upper bounds on the possible mass of the
lightest Higgs boson are
\begin{eqnarray}
\mhl\lsim 130\gev && {\rm MSSM}, \\
\mhl\lsim 150\gev && {\rm any\ SUSY\ GUT}, \\
\mhl\lsim 200\gev && {\rm any\ model\ with}\\
&&\rm GUT\ and\ desert.\nonumber
\end{eqnarray}
In the largest part of parameter space, e.g. $\mha>150$ GeV in the MSSM,
the lightest Higgs boson has fairly SM-like couplings.

The first discovery of a light Higgs boson is likely to occur at the LHC which
might be operating for several years before a next-generation lepton collider
is built. Following its discovery, interest will focus on measurements of its
mass, total width, and partial widths.
A first question then is what could be accomplished at the Large Hadron
Collider (LHC) or the Next Linear Collider (NLC) in this regard.

\begin{figure}
\let\normalsize=\captsize   %%%% changes the font to "\small"
\begin{center}
\centerline{\psfig{file=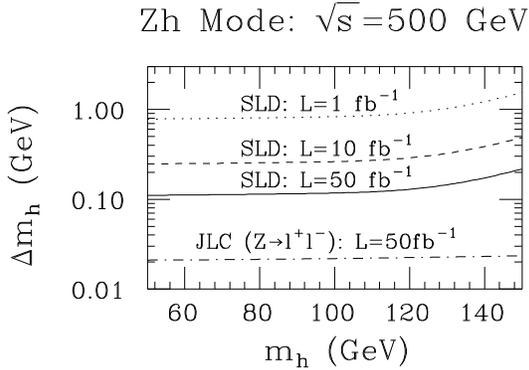,width=7cm}}
\begin{minipage}{7cm}       %%%% reduces width of caption to 7cm
\smallskip
\caption{{\baselineskip=0pt
The uncertainty $\pm\Delta\mh$ in the
determination of $\mh$ for a SM-like Higgs boson using $Z\h$ production
and a $\pm 4\gev$ (``SLD'') or $\pm0.3\gev$ (``JLC'')
single event mass resolution for $\mh$.}}
\label{massresolution}
\end{minipage}
\end{center}
\vspace*{-.3in}
\end{figure}

At the LHC, a SM-like
Higgs can be discovered either through gluon fusion, followed by
$\gamma\gamma$ or $4\ell$ decay,
\begin{eqnarray}
&& gg\to h \to \gamma \gamma\;, \\
&& gg\to h \to Z\zstar \to 4l\;,
\end{eqnarray}
or through associated production
\begin{eqnarray}
&& gg \to t\overline{t}h\nonumber\\ \noalign{\vskip-.7ex}
&& \hspace{3.5em}\raise1ex\hbox{$\vert$}\hspace{-.5em}\to \gamma \gamma\;, \\
&& q\overline{q}\to Wh\nonumber\\ \noalign{\vskip-.7ex}
&& \hspace{3.75em}\raise1ex\hbox{$\vert$}\hspace{-.5em}\to \gamma \gamma\;.
\end{eqnarray}
The LHC collaborations report that the Higgs boson is
detectable in the mass range $50\lsim m_h\lsim 150\gev$ via its
$\gamma \gamma$ decay mode. The mass resolution is expected to be $\lsim 1\%$.
At the NLC the Higgs boson is produced in the Bjorken process
\begin{equation}
\ee \to \zstar  \to Z\h 
%\nonumber\\ \noalign{\vskip-.3ex}
%&& \hspace{7.5em}\raise1ex\hbox{$\vert$}\hspace{-.5em}\to Zb\bar b\;,
\end{equation}
and the $\h$ can be studied through its dominant $b\bar b$ decay.
At the NLC (which may be available prior to 
a $\mm$ collider) the mass resolution is strongly dependent on the detector
performance and signal statistics:
\begin{equation}
\Delta \mh^{} \simeq R_{\rm event}({\rm GeV})/\sqrt N \;,
\end{equation}
where $R_{\rm event}$ is the single event
resolution and $N$ is the number of signal
events. The single event resolution is about $4\gev$ for an
SLD-type detector\cite{janot},
but improved performance as typified by the ``super''-LC detector would make
this resolution about $0.3\gev$\cite{jlci,kawagoe}. The uncertainty in the
Higgs boson mass for various integrated luminosities is shown
in Fig.~\ref{massresolution}.
For a Higgs boson with Standard
Model couplings this gives a Higgs mass determination of
\begin{eqnarray}
\Delta \mhsm\simeq 400\mev\left ({{10\fbi}\over {L}}\right )^{1/2}\;,
\end{eqnarray}
for the SLD-type detector.

%2

Precision measurements of the
Higgs total width and partial widths will be
necessary to distinguish between the predictions of the SM Higgs boson $\hsm$
and the MSSM Higgs boson $\hl$.
Can the total and partial widths be measured at other machines? This is a
complicated question since each machine contributes different pieces to the
puzzle. The bottom line\cite{gsw} 
is that the LHC, NLC, and $\gamma \gamma$ colliders
each measure interesting couplings and/or branching ratios, 
but their ability to detect deviations due
to the differences between the $\hl$ and $\hsm$ is limited to $\mha\lsim
300\gev$.  Further,
a model-independent study of all couplings and widths requires all three
machines with consequent error propagation problems.

\begin{figure}[t]
\let\normalsize=\captsize   %%%% changes the font to "\small"
\begin{center}
\centerline{\psfig{file=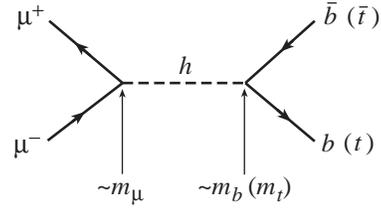,width=5cm}}
\begin{minipage}{7cm}       %%%% reduces width of caption to 7cm
\smallskip
\caption{{\baselineskip=0pt
Feynman diagram for $s$-channel production of a Higgs boson.}}
\label{schanfig}
\end{minipage}
\end{center}
\vspace*{-.25in}
\end{figure}

The $s$-channel process $\mm \to b\overline{b}$
shown in Fig.~\ref{schanfig} is
uniquely suited to several critical precision  Higgs boson 
measurements \cite{mupmumprl,bbgh}.
 Detecting and studying the
Higgs boson in the  $s$-channel would require that the machine energy be
adjusted to correspond to the Higgs mass.
 Since the storage ring is only a
modest fraction of the overall muon
collider cost\cite{palmer2}, a special-purpose ring could be built to
optimize the luminosity near the Higgs peak.

\begin{figure}
\let\normalsize=\captsize   %%%% changes the font to "\small"
\begin{center}
\centerline{\psfig{file=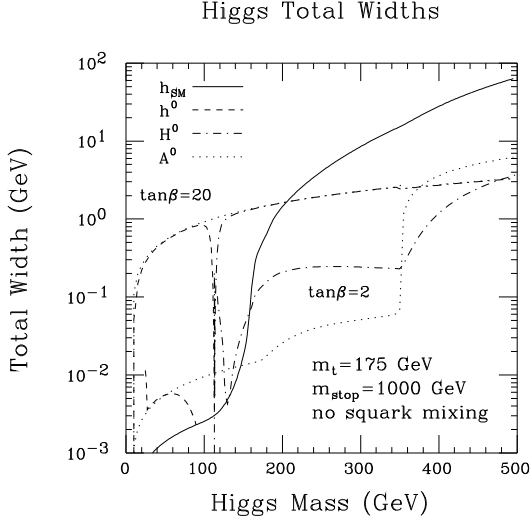,width=7cm}}
\begin{minipage}{7cm}       %%%% reduces width of caption to 7cm
\smallskip
\caption{{\baselineskip=0pt
Total width versus mass of the SM and MSSM Higgs bosons
for $\mt=175\gev$.
In the case of the MSSM, we have plotted results for
$\tan\beta =2$ and 20, taking $\mstop=1\tev$ and
including two-loop corrections following
Refs.~\protect\cite{habertwoloop,carenatwoloop}
neglecting squark mixing; SUSY decay channels are assumed to be absent.}}
\label{hwidths}
\end{minipage}
\end{center}
\vspace*{-.2in}
\end{figure}

The $s$-channel Higgs phenomenology is set
by the $\sqrt{s}$ rms Gaussian spread
denoted by $\sigrts$.
A convenient formula for $\sigrts$ is
\begin{equation}
\sigrts = (7~{\rm MeV})\left({R\over 0.01\%}\right)\left({\rts\over {\rm
100\ GeV}}\right) \ .
\label{resolution}
\end{equation}
A crucial consideration is how this natural spread in the muon collider
beam energy compares to the width of the Higgs bosons, given in
Fig.~\ref{hwidths}. In particular, a direct scan measurement
of the Higgs width requires a beam spread
comparable to the width. The narrowest Higgs boson widths are those
of a light SM Higgs boson with mass $\lsim 100\gev$.
In the limit where the heavier MSSM Higgs bosons
become very massive, the lightest supersymmetric Higgs 
typically has a mass of order 100 GeV and has couplings that are sufficiently
SM-like that its width approaches that of a light $\hsm$ of the same mass.
In either case, the discriminating power of a 
muon collider with a very sharp energy resolution would
be essential for a direct width measurement.

%3

%4

A quantitative examination of Fig.~\ref{hwidths} shows that 
for typical muon beam resolution ($R=0.06\%$)
\begin{eqnarray}
\sigrts &\gg& \Gamma _{h_{SM}}\;, \ {\rm for\ } \mhsm\sim 100\gev\;,\\
\sigrts &\sim& \Gamma _{\hl}\;, \ {\rm for\ } \mhl \ {\rm not \ near} \ 
\mhl^{\rm max}\;,\\
\sigrts &\lsim& \Gamma _{\hh},\Gamma _{\ha}\;, \ 
{\rm at \ moderate\ } \tan \beta\;,\\
&& \ {\rm for\ } m_{\hh,\ha}\sim400\gev\;,\nonumber\\
&\ll& \Gamma _{\hh},\Gamma _{\ha}\;, \ {\rm at \ large}\ \tan \beta\;,\\
&& \ {\rm for\ } m_{\hh,\ha}\sim400\gev\;.\nonumber
\end{eqnarray}
To be sensitive to the $\Gamma_{\hsm}$ case, a resolution 
$R\sim0.01\%$ is mandatory. This is an important conclusion given that
such a small resolution requires early consideration in the machine design.

\begin{figure}
\let\normalsize=\captsize   %%%% changes the font to "\small"
\begin{center}
\centerline{\psfig{file=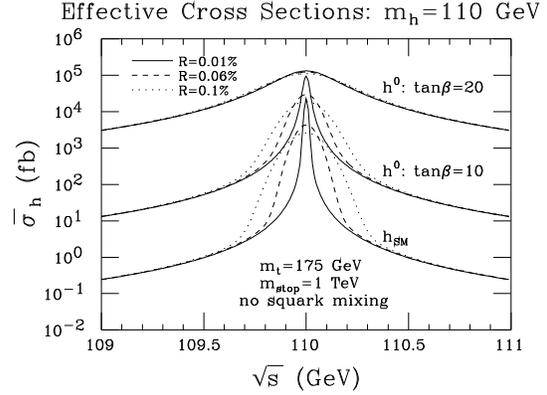,width=7cm}}
\begin{minipage}{7cm}       %%%% reduces width of caption to 7cm
\caption{The effective cross section, $\sighbar$,
obtained after convoluting $\sigma_{\h}$ with
the Gaussian distributions for $R=0.01\%$, $R=0.06\%$, and $R=0.1\%$, is
plotted as a function of $\protect\rts$ taking $\mh=110\gev$.
}
\label{gausssigma}
\end{minipage}
\end{center}
\vspace*{-.2in}
\end{figure}

The $s$-channel Higgs resonance cross section is
\begin{equation}
\sigma_{\h} = {4\pi \Gamma(\h\to\mu\mu) \, \Gamma(\h\to X)\over
\left(\hat s-\mh^2\right)^2 + \mh^2 [\gamh]^2} \;,
\label{basicsigma}
\end{equation}
where $\shat =(p_{\mu^+}+p_{\mu^-})^2$ is the c.~m.\ energy
squared of the event,
$X$ denotes a final state and $\gamh$ is the total width.
The effective cross section is obtained by convoluting this resonance
form with the Gaussian distribution of width $\sigrts$ centered at $\rts$.
When the Higgs width is much smaller than $\sigrts$,
the effective signal cross section result for $\rts=\mh$,
denoted by $\sighbar$, is
\begin{equation}
\sighbar
={2\pi^2 \Gamma(\h\to\mu\mu)\, BF(\h\to X) \over \mh^2}\times
{1\over \sigrts \sqrt{2\pi}}\;.
\label{narrowwidthsigma}
\end{equation}
In the other extreme, where the Higgs width is much broader than
$\sigrts$\,, at $\rts=\mh$ we obtain
\begin{equation}
\sighbar={4\pi BF(\h\to\mu\mu)BF(\h\to X)\over \mh^2}\;.
\label{broadwidthsigma}
\end{equation}
Figure~\ref{gausssigma}
illustrates the result of this convolution as a function of $\rts$
for $\rts$ near $\mh$ in the three situations: $\gamh\ll\sigrts$,
$\gamh\sim\sigrts$ and $\gamh\gg\sigrts$. We observe that small $R$
greatly enhances 
the peak cross section for $\rts=\mh$ when $\gamh\ll\sigrts$,
as well as providing an opportunity to directly measure $\gamh$.

%5

As an illustration,
suppose $\mh\sim 110$ GeV and $\h$ is detected in $e^+e^-\to Z\h$ or
$\mu^+\mu^-\to Z\h$ with mass uncertainty
$\delta \mh\sim \pm 0.8$ GeV (obtained with luminosity $L\sim 1 \fbi$).
For a standard model Higgs of this mass, the
width is about 3.1 MeV. How many scan points and how much luminosity
are required to zero in on
$\mhsm$ to within one rms spread $\sigrts$?
For $R=0.01\%$ ($R=0.06\%$), $\sigrts\sim 7.7\mev$ ($\sim 45\mev$)
and the number of scan points required to cover the $1.6\gev$
mass zone at intervals of $\sigrts$ will be 230 (34), respectively.
The luminosity required to observe (or exclude) the Higgs at each point
is $L\gsim 0.01\fbi$ ($L\gsim 0.3\fbi$) for $R=0.01\%$ ($R=0.06\%$).
Thus, the total luminosity required to zero in on the Higgs will
be $\sim 2.3\fbi$ ($\sim 10.2\fbi$) in the two cases.

More generally, the $L$ required at each scan point
decreases as (roughly) $R^{1.7}$, whereas the number
of scan points only grows like $1/R$, implying that
the total $L$ required for the scan decreases as $\sim R^{0.7}$.
Thus, the $\mm$ collider should be constructed with the smallest
possible $R$ value with the proviso that the number
of $\rts$ settings can be correspondingly increased for the required scan.
It must be possible
to quickly and precisely adjust the energy of the $\mm$ collider
to do the scan.

%If we simply pursue this scan, and then fill in every 2 MeV once $m_h$ is
%roughly located we end up with: some reasonable
%determination of $\Gamma _h$ (and $m_h$
%to within a fraction of $\Gamma _h$) to about 10\%.

To measure the width of a SM-like Higgs boson, 
one would first determine $\mh$ to within $d\sigrts $
with $d\lsim 0.3$ and then measure the cross section accurately at the wings of
the excitation peak, see Fig.~\ref{gausssigma}.
The two independent measurements of $\sigma_{\rm
wings}/\sigma_{\rm peak}$ give improved precision for the Higgs mass and
determine the Higgs width.
It is advantageous to put more luminosity on
the wings than the peak. Thus, to extract the total width we propose
the following procedure\cite{bbgh}.
First, conduct a rough scan to determine $\mh$
to a precision $\sigrts d$, with $d\lsim 0.3$.
Then perform three measurements.
At $\rts_1=\mh+\sigrts d$
expend a luminosity $L_1$ and
measure the total rate $N_1=S_1+B_1$.
Then perform measurements at
\begin{equation}
 \rts_2=\rts_1-\nsigrts\sigrts \label{measure1}
\end{equation}
 and one at
\begin{equation}
\rts_3=\rts_1+\nsigrts\sigrts \label{measure2}
\end{equation}
yielding $N_2=S_2+B_2$ and $N_3=S_3+B_3$
events, respectively, with luminosities of
$L_2=\rho_2L_1$ and $L_3=\rho_3L_1$. The backgrounds can be determined from
measurements farther from the resonance or from theoretical predictions.
Next evaluate the ratios $r_2=(S_2/\rho_2)/S_1 $ and
$r_3=(S_3/\rho_3)/S_1 $, for which the partial decay  rates in the
numerator in Eq.~(\ref{basicsigma}) cancel out. Since the excitation curve
has a specific shape given by convoluting 
the denominator in Eq.~(\ref{basicsigma}) with the Gaussian distribution,
these measured ratios determine the mass
and total width of the Higgs boson.
We find that the choices $\nsigrts\simeq2$ and $\rho_2=\rho_3\simeq2.5$ 
are roughly optimal when $\sigrts\gsim\gamh$.
For these choices and $R=0.01\%$, a total luminosity $L=L_1+L_2+L_3$
of $2\fbi$ ($200\fbi$) would be required to measure $\gamh$ with
an accuracy of $\pm30\%$ for $\mh=110\gev$ ($\mh=\mz$).
An accuracy of $\pm 10\%$ for $\gamh$ could be achieved for reasonable
luminosities provided $\mh$ is not near $\mz$.

It must be stressed that the ability to precisely determine the
energy of the machine when the three measurements are taken is crucial
for the success of the three-point technique. A mis-determination of the
{\it spacing} of the measurements in Eqs.~(\ref{measure1}) and (\ref{measure2})
by just 3\% would result in an error in $\gamhsm$ of 30\%. This does not
present a problem provided some polarization of the beam can be achieved
so that the precession of the spin of the muon as it circulates in the final
storage ring can be measured. Given this and
the rotation rate, the energy can be determined
to the nearly 1 part in a million accuracy required. This energy calibration
capability must be incorporated in the machine design from the beginning.

The other quantity that can be measured with great precision at
a $\mm$ collider for a SM-like Higgs with $\mh\lsim 130\gev$
is $G(b\anti b)\equiv\Gamma(\h\to\mm)BF(\h\to b\anti b)$. 
For $L=50\fbi$ and $R=0.01\%,0.06\%$, $G(b\anti b)$ can be measured
with an accuracy of $\pm0.4\%,\pm2\%$ ($\pm3\%,\pm15\%$) at $\mh=110\gev$
($\mh=\mz$). By combining this measurement with the $\pm\sim 7\%$
determination
of 
\break
$BF(\h\to b\anti b)$ that could be made in the $Z\h$ production mode,
a roughly $\pm 8-10\%$ determination of $\Gamma(\h\to\mm)$ becomes
possible. ($R=0.01\%$ is required if $\mh\sim\mz$.)

Suppose we find a light Higgs $h$ and measure its mass, total width and partial
widths. The critical questions that then arise are:
\begin{itemize}
\item Can we determine if the particle is a SM Higgs or a
supersymmetric Higgs?
\item If the particle is a supersymmetric Higgs
 boson, say in the MSSM,
can we then predict masses of the heavier Higgs bosons $\hh$, $\ha$, and
$H^\pm$ in order to discover them in subsequent measurements?
\end{itemize}
In the context of the MSSM, the answers to these questions can
be delineated. 

Enhancements of $\gamh$ of order 30\% 
relative to the prediction for the SM $\hsm$ are the
norm (even neglecting possible SUSY decays) for $\mha\lsim\break
 400\gev$.
A 10\% measurement of $\gamh$ would thus be relatively likely
to reveal a $3\sigma$ statistical enhancement.
However, using
the deviation to determine the value of $\mha$
is model-dependent. For example, if $\mh=110\gev$
and there is no stop mixing, then the percentage deviation would
fairly uniquely fix $\mha$, whereas if $\mh=110\gev$ and there
is maximal stop mixing, as defined in Ref.~\cite{gsw}, then
the measured deviation would only imply a relation
between $\tanb$ and $\mha$.

$\gamh$ could be combined with branching ratios
to yield a more definitive determination of $\mha$. For instance,
we can compute $\Gamma(\h\to b\anti b)=\gamh BF(\h\to b\anti b)$
using $BF(\h\to b\anti b)$ as measured in $Z\h$ production.
It turns out that the percentage deviation of this partial width 
for the $\hl$ from the $\hsm$ prediction 
is rather independent of $\tanb$ and gives a mixing-independent
determination of $\mha$, which, after including systematic
uncertainties in our knowledge of $\mb$,
would discriminate between a value of $\mha\leq 300\gev$ vs. $\mha=\infty$
at the $\geq3\sigma$ statistical level.

Returning to $\Gamma(\h\to \mm)$,
deviations at the $\gsim 3\sigma$ statistical level
in the prediction for this partial width for the $\hl$ as compared
to the $\hsm$ are predicted out to $\mha\gsim 400\gev$.
Further, the percentage of deviation from the SM prediction
would provide a relatively
accurate determination of $\mha$ for $\mha\lsim 400\gev$.
For example, if $\mh=110\gev$,
$\Gamma(\hl\to\mm)$ changes by $20\%$ (a $\gsim 2\sigma$ effect)
as $\mha$ is changed from $300\gev$ to $365\gev$.

Deviations for other quantities, e.g. $BF(\h\to b\anti b)$,
depend upon the details of the stop squark masses and mixings,
the presence of SUSY decay modes, and so forth,
much as described in the case of $\gamh$.  Only partial widths
provide a mixing-independent determination of $\mha$.
The $\mm$ collider provides, as described, as least two particularly unique
opportunities for determining two very important partial
widths, $\Gamma(\h\to b\anti b)$ and $\Gamma(\h\to \mm)$,
thereby allowing a test of the predicted proportionality
of these partial widths to fermion mass independent of the lepton/quark
nature of the fermion.

Thus, if $\mha\lsim 400\gev$, 
we may gain some knowledge of $\mha$ through precision
measurements of the $\hl$'s partial widths.  This
would greatly facilitate direct observation of the $\ha$
and $\hh$ via $s$-channel production at a $\mm$ collider
with $\rts\lsim 500\gev$.  As discussed in more detail shortly,
even without such pre-knowledge of $\mha$, discovery of the $\ha,\hh$
Higgs bosons would be possible in the $s$-channel at a $\mm$
collider provided that $\tanb\gsim 3-4$.  With pre-knowledge
of $\mha$, detection becomes possible for $\tanb$ values not far
above 1, provided $R\sim 0.01\%$ (crucial since the $\ha$ and $\hh$
become relatively narrow for low $\tanb$ values).

Other colliders offer various mechanisms
to directly search for the $\ha,\hh$, but also have limitations:
\begin{itemize}
\item The LHC has a discovery hole and ``$\hl$-only'' regions at moderate
$\tan \beta$, $\mha\gsim 200\gev$.
\item At the NLC one can use the mode $\ee\to \zstar\to \hh\ha$ 
(the mode $\hl\ha$ is suppressed for
large $\mha$), but it is limited to $\mhh\sim \mha\lsim \sqrt{s}/2$.
\item A $\gamma \gamma$ collider could probe heavy Higgs up to masses of
$\mhh\sim \mha\sim 0.8\sqrt{s}$, but this would quite likely require
$L\sim 100{\fb}^{-1}$, especially if the Higgs bosons are at the upper
end of the  $\gamma \gamma$ collider energy spectrum\cite{ghgamgam}.
\end{itemize}

Most GUT models predict $\mha\gsim 200\gev$, and perhaps as large
as a TeV\cite{gut}. For large $\mha\sim\mhh$, $s$-channel searches can be made
at a $\mm$ collider up to $\sim\rts$, whereas the $\zstar\to\hh\ha$ mode at an
$\ee$ collider fails for $\mha\sim\mhh\gsim \rts/2$.
In particular, at a muon collider with $\rts\sim 500\gev$, scan detection
of the $\ha,\hh$ is possible in the mass range from 200 to 500 GeV in
$s$-channel production, provided $\tanb\gsim 3-4$,
whereas an $\ee$ collider of the same energy can only probe $\mhh\sim\mha\lsim
220\gev$. That the signals become viable when $\tanb>1$
(as favored by GUT models) is due to the fact that the couplings
of $\ha$ and (once $\mha\gsim 150\gev$) 
$\hh$ to $b\overline{b}$ and, especially to
$\mu^+\mu^-$,  are
proportional to $\tanb$, and thus increasingly enhanced as
$\tanb$ rises.

Although the $\mm$ collider cannot discover the $\hh,\ha$
in the $\tanb\lsim 3$ region, this
is a range in which the LHC {\it could} find
the heavy Higgs bosons in a number of modes. 
That the LHC and the NMC are complementary in this
respect is a very crucial point. Together, discovery of the $\ha,\hh$
is essentially guaranteed.

If the $\hh,\ha$ are observed at
the $\mm$ collider, measurement of their widths 
will typically be straightforward.  For moderate $\tanb$ the $\ha$ and $\hh$
resonance peaks do not overlap
and $R\lsim 0.06\%$ will be adequate, since for such $R$ values
$\Gamma _{\hh,\ha}\gsim \sigrts$. However, if $\tanb$ is large,
then for most of the $\mha\gsim 200\gev$ parameter range
the $\ha$ and $\hh$ are sufficiently degenerate that
there is significant overlap of the $\ha$ and $\hh$ resonance peaks.
In this case, $R\sim 0.01\%$ resolution would be
necessary for observing the double-peaked structure and
separating the $\ha$ and $\hh$ resonances.

A $\sqrt{s}\sim 500\gev$ muon collider still might not have sufficient energy
to discover heavy supersymmetric Higgs bosons. Further, distinguishing
the MSSM from the SM by detecting small deviations
of the $\hl$ properties from those predicted for the $\hsm$ becomes quite
difficult for $\mha\gsim400\gev$. However, construction of a higher energy
machine, say $\sqrt{s}=4\tev$, 
would allow discovery of $\ha,\hh$ in the $b\anti b$ or $t\anti t$
channels (see the discussion in Section 5).

\begin{figure}[h]
\let\normalsize=\captsize   %%%% changes the font to "\small"
\begin{center}
\centerline{\psfig{file=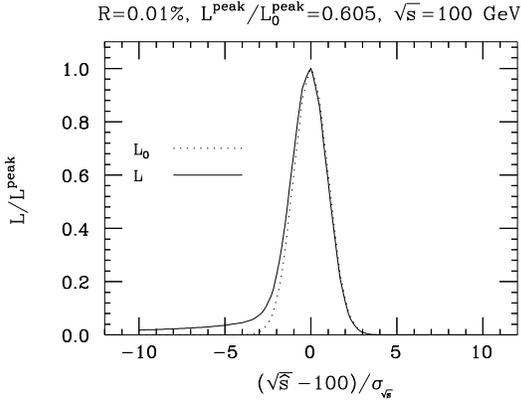,width=7cm}}
\begin{minipage}{7cm}       %%%% reduces width of caption to 7cm
\smallskip
\caption{{\baselineskip=0pt
$d{\cal L}/d\protect\rtshat$ relative to its peak value at
$\protect\rtshat=\protect\rts$
is plotted before and after soft-photon radiation.
We have taken $\protect\rts=100\gev$ and  $R=0.01\%$. The ratio
of peak height after including soft-photon radiation to
that before is 0.605.}}
\label{bremdistfig}
\end{minipage}
\end{center}
\vspace*{-.05in}
\end{figure}

We close this section with brief comments on the effects of bremsstrahlung
and beam polarization. Soft photon
radiation must be included when determining the resolution in energy and the
peak luminosity achievable at an $\ee$ or $\mm$ collider.
This radiation is substantially reduced at a $\mm$ collider due to the
increased mass of the muon compared to the electron. In
Fig.~\ref{bremdistfig} we show the luminosity distribution
before and after including the soft photon radiation. These bremsstrahlung
effects are calculated in
Ref.~\cite{bbgh}. A long tail extends down
to low values of the energy.

%6

\begin{figure}[t]
%8
%\begin{figure}
\let\normalsize=\captsize   %%%% changes the font to "\small"
\begin{center}
\centerline{\psfig{file=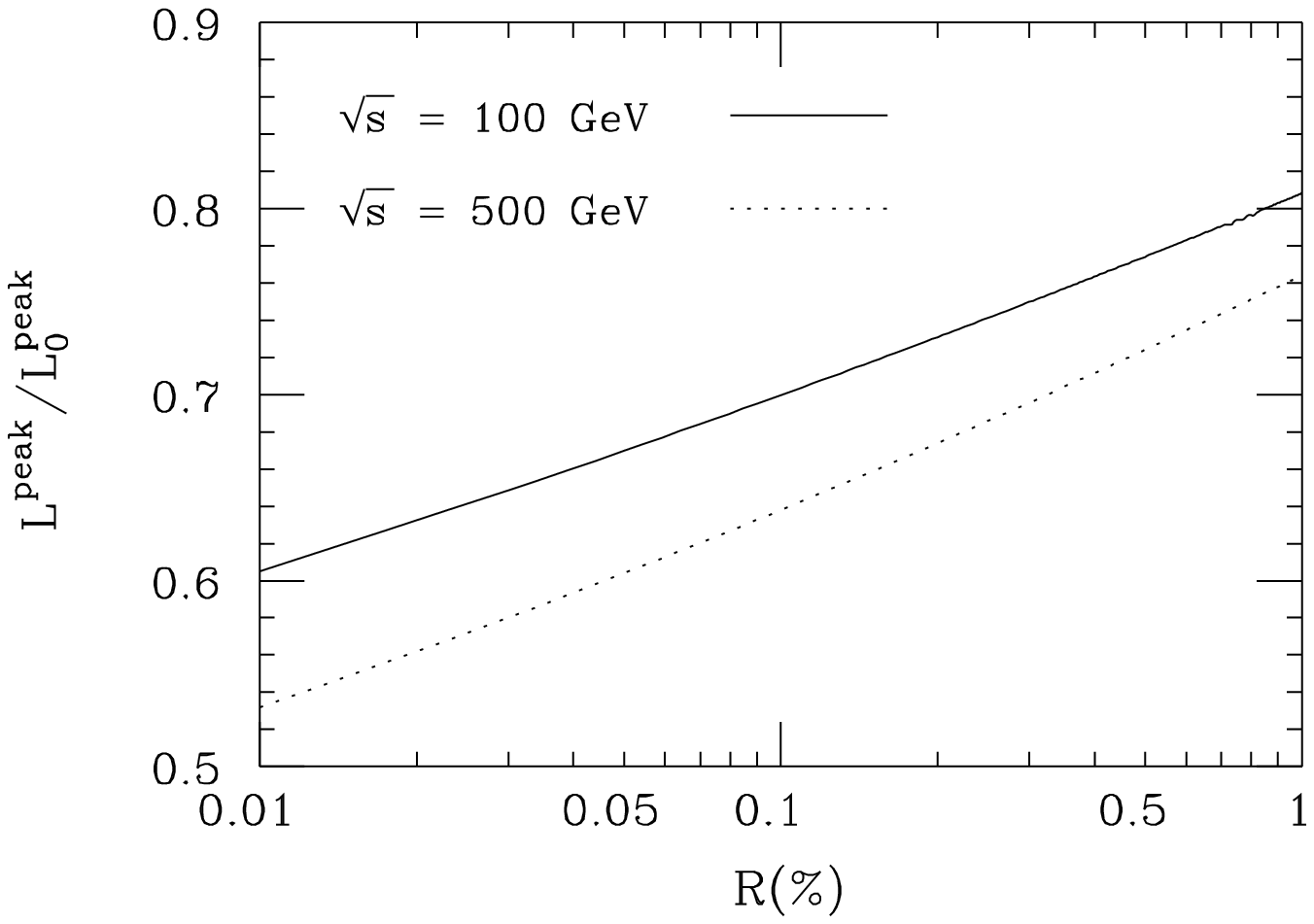,width=6.75cm}}
\begin{minipage}{7cm}       %%%% reduces width of caption to 7cm\smallskip
\smallskip
\caption{{\baselineskip=0pt
$\left.{d{\cal L}\over d\protect\rtshat}/
{d{\cal L}_0\over d\protect\rtshat}\right|_{\protect\rtshat=\protect\rts}$
as a function of $R$ for $\protect\rts=100$ and 500 GeV.}}
\label{brempeakfig}
\end{minipage}
\end{center}
\vspace{-.2in}
\end{figure}

For a SM-like Higgs boson with width smaller than $\sigrts$, the primary
effect of bremsstrahlung is a reduction in the peak luminosity.
The ratio of the luminosity  peak height after and before including the
bremsstrahlung is shown in Fig.~\ref{brempeakfig}. The conclusions above
regarding $s$-channel Higgs detection 
are those obtained with inclusion of\break
bremsstrahlung effects.

%\bigskip

\begin{figure}[h]
\let\normalsize=\captsize   %%%% changes the font to "\small"
\begin{center}
\centerline{\psfig{file=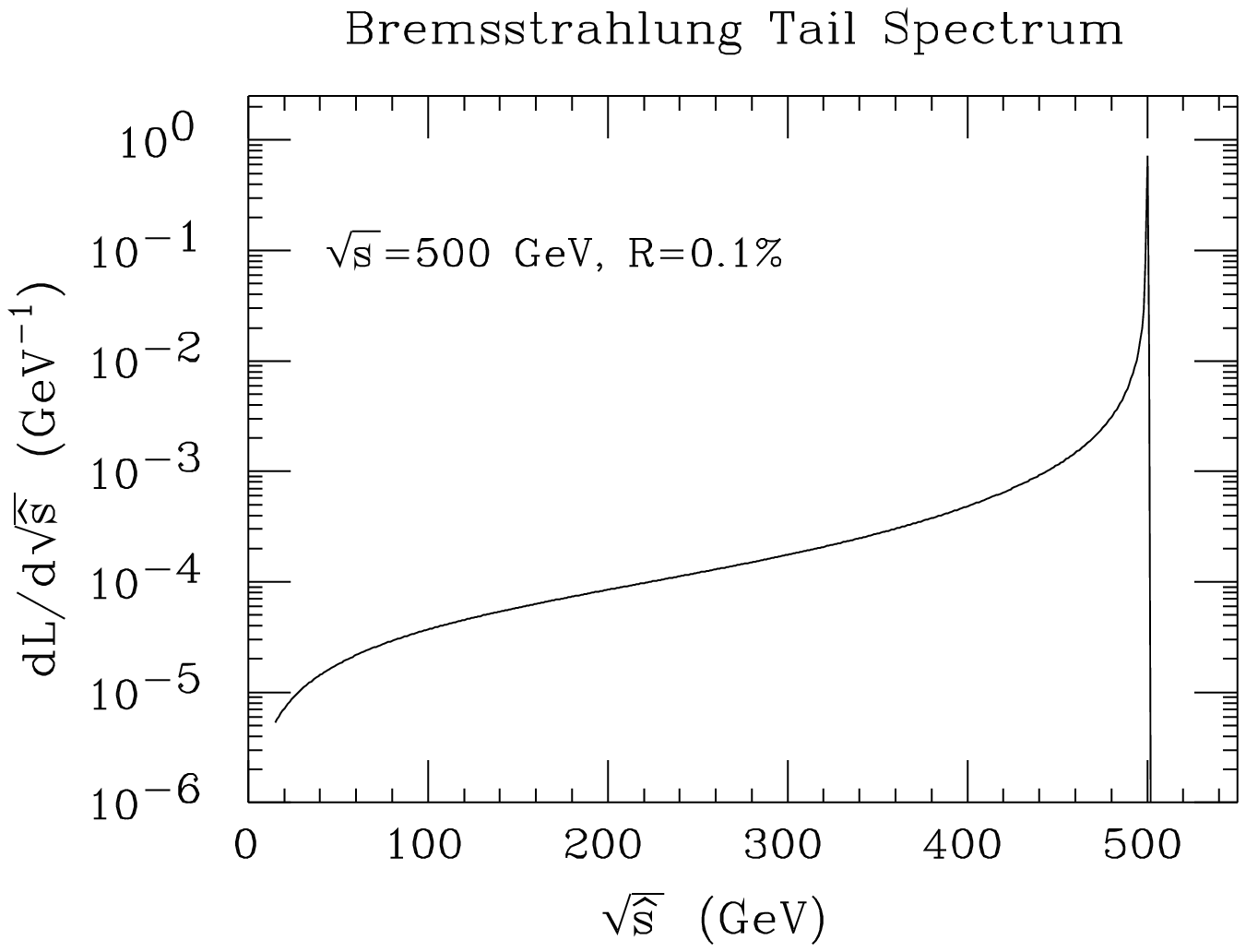,width=6.75cm}}
\begin{minipage}{7cm}       %%%% reduces width of caption to 7cm
\smallskip
\caption{{\baselineskip=0pt
${d{\cal L}\over d\protect\rtshat}$ as a function of
$\protect\rtshat$
for $R=0.1\%$ and $\protect\rts=500\gev$. The integral under the
curve is normalized to 1.}}
\label{bremtail}
\end{minipage}
\end{center}
%\end{figure}
\end{figure}

The low-energy bremsstrahlung tail provides a
self-scan over the range of energies below 
the design energy, and thus can 
be used to detect $s$-channel resonances. The full luminosity
distribution for the tail is shown in Fig.~\ref{bremtail}.
Observation of $\ha,\hh$ peaks in the $b\anti b$ mass distribution
$m_{b\anti b}$ created by this bremsstrahlung tail may be possible.
The region of the $(\mha,\tanb)$ parameter space plane for which
a peak is observable depends strongly on the $b\anti b$
invariant mass resolution. For an excellent $m_{b\anti b}$
mass resolution of order
$\pm 5\gev$ and integrated luminosity
of $L=50\fbi$ at $\rts=500\gev$, 
the $\ha,\hh$ peak(s) are observable for $\tanb\gsim 5$
at $\mha\gsim 400\gev$ (but only for very large $\tanb$ values
in the $\mha\sim \mz$ region due to the large $s$-channel $Z$
contribution to the $b\anti b$ background).

%7

In the $s$-channel Higgs studies, polarization of the muon beams could present
a significant advantage over the unpolarized case, since signal and background
come predominantly from different polarization states. Polarization $P$
of both beams would enhance the significance of a Higgs signal
provided the factor by which the luminosity is reduced
is not larger than $(1+P^2)^2/(1-P^2)$.
For example, a reduction in luminosity by a factor of 10
could be compensated by a polarization $P=0.84$, leaving the significance of
the signal unchanged\cite{parsa}. Furthermore, {\it transverse} polarization of
the muon
beams could prove useful for studying CP-violation in the Higgs sector.
Muons are produced naturally polarized from $\pi$ and $K$ decays. An important
consideration for the future design of muon colliders is the extent to which
polarization can be maintained through the cooling and acceleration processes.

\section{\uppercase{Precision threshold studies}}

Good beam energy resolution
is crucial for the determination of the Higgs width. 
Another area of physics where the naturally 
good resolution of a $\mm$ collider would prove valuable is
studies of the $t\overline{t}$ and $W^+W^-$ thresholds, 
similar to those proposed for the NLC and LEP~II. 
The $t\overline{t}$ threshold shape 
determines $\mt$, $\Gamma _t$ and
the strong coupling $\alpha _s$, while the $W^+W^-$ threshold shape 
determines $\mw$ and possibly also $\Gamma_W$.
At a $\mm$ collider, even a conservative natural beam resolution 
$R\sim 0.1\%$ would allow substantially increased precision in the measurement
of most of these quantities as compared to other machines.
Not only is such monochromaticity already greatly superior
to $\ee$ collider designs, where typically $R\sim1\%$,
but also at a $\mm$ collider there is no significant beamstrahlung 
and the amount of initial state radiation (ISR) is greatly reduced.
ISR and, especially, beam smearing cause significant loss of precision in
the measurement of the top quark and $W$ masses at $\ee$ colliders.

\begin{figure}[t]
\let\normalsize=\captsize   %%%% changes the font to "\small"
\begin{center}
\centerline{\psfig{file=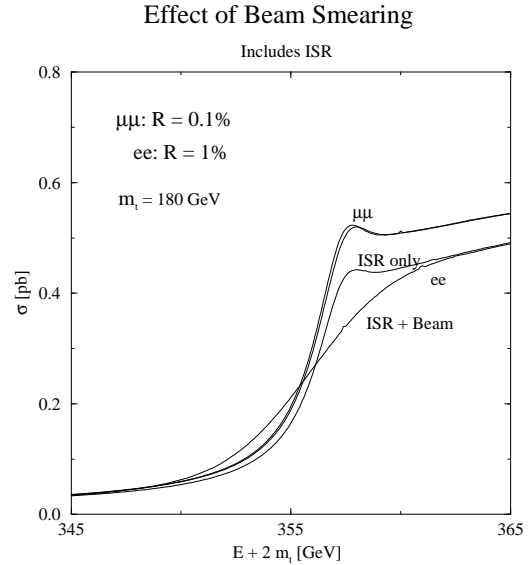,width=7cm}}
\begin{minipage}{7cm}       %%%% reduces width of caption to 7cm
\bigskip
\caption{{\baselineskip=0pt
The threshold curves are shown for
$\mm$ and $\ee$ machines including ISR and with and without
beam smearing. Beam smearing has only a small effect
at a muon collider, whereas at an electron collider the threshold region is
significantly smeared. The strong coupling is taken
to be $\alpha_s(\mz)=0.12$.}}
\label{ttbarfig}
\end{minipage}
\end{center}
\vspace*{-.3in}
\end{figure}

To illustrate, consider 
threshold production of the top quark, which has been extensively studied for
$\ee$ colliders\cite{ttbaree}. 
Figure~\ref{ttbarfig} shows the effects of including beam smearing and ISR
for the threshold production of top quarks using a Gaussian beam spread of
$1\%$ for the $\ee$ collider\cite{ttbar}. Also shown are our corresponding
results for the $\mm$ collider with $R=0.1\%$, see \cite{ttbar}. 
The threshold peak is no longer washed out in the $\mm$ case.
The precision with which one could measure $\mt$, $\alpha_s$
and $\Gamma_t$ at various facilities is shown in Table~\ref{tablei}.
Improvements in the determination of $\mw$ should also be
possible\cite{dawson}.

%9

\begin{table*}
\centering
\caption[]{\label{tablei}\small
Measurements of the standard model parameters: top mass $\mt$, strong coupling
$\alpha_s$, and top quark width $\Gamma_t$.}
\medskip
\begin{tabular}{|l|c|c|c|c|}
\hline
\multicolumn{1}{|c|}{}
&\multicolumn{1}{|c|}{Tevatron}
&\multicolumn{1}{|c|}{LHC}
&\multicolumn{1}{|c|}{NLC}
&\multicolumn{1}{|c|}{FMC}
\\
\multicolumn{1}{|c|}{}
&\multicolumn{1}{|c|}{(1000 $pb^{-1}$)}
&\multicolumn{1}{|c|}{(20 $pb^{-1}$)}
&\multicolumn{1}{|c|}{(10 $fb^{-1}$)}
&\multicolumn{1}{|c|}{(10 $fb^{-1}$)}
\\
\multicolumn{1}{|c|}{}
&\multicolumn{1}{|c|}{(10 $fb^{-1}$)}
&\multicolumn{1}{|c|}{}
&\multicolumn{1}{|c|}{}
&\multicolumn{1}{|c|}{}
\\ \hline \hline
\multicolumn{1}{|c|}{$\Delta \mt ({\rm GeV})$}
&\multicolumn{1}{|c|}{$ 4$}
&\multicolumn{1}{|c|}{2}
&\multicolumn{1}{|c|}{$0.52\cite{igo}$}
&\multicolumn{1}{|c|}{$0.3$}
\\
\multicolumn{1}{|c|}{}
&\multicolumn{1}{|c|}{$ 1$}
&\multicolumn{1}{|c|}{}
&\multicolumn{1}{|c|}{}
&\multicolumn{1}{|c|}{}
\\ \hline
\multicolumn{1}{|c|}{$\Delta \alpha _s$}
&\multicolumn{1}{|c|}{}
&\multicolumn{1}{|c|}{}
&\multicolumn{1}{|c|}{0.009}
&\multicolumn{1}{|c|}{0.008}
\\ \hline
\multicolumn{1}{|c|}{$\Delta \Gamma _t/\Gamma _t$}
&\multicolumn{1}{|c|}{$ 0.3\cite{yuan}$}
&\multicolumn{1}{|c|}{}
&\multicolumn{1}{|c|}{0.2}
&\multicolumn{1}{|c|}{{\rm better}}
\\ \hline
\end{tabular}
\end{table*}

The value of such improvements in precision can be substantial.
Consider precision electroweak corrections, for example.
The prediction for 
the SM or SM-like Higgs mass $\mh$ depends on $\mw$ and $\mt$
through the one-loop equation
\begin{eqnarray}
\mw^2&=&\mz^2
\left [1-{{\pi\alpha}\over {\sqrt{2}G_\mu \mw^2(1-\delta r)}}\right ]^{1/2}
\!\!\!\!\!\!, \qquad
\end{eqnarray}
where $\delta r$ depends quadratically on $\mt$ and logarithmically on $\mh$.
Current expectations for LEP~II and the Tevatron imply precisions of order
\begin{eqnarray}
\Delta \mw&=&40\mev\;, \\
\Delta \mt&=&4\gev\;.
\label{precision}
\end{eqnarray}
For the uncertainties of Eq.~(\ref{precision})
and the current central values of $\mw=80.4$~GeV and $\mt=180$~GeV, the Higgs
mass would be constrained to the $1\sigma$ range
\begin{equation}
50<\mh<200 \gev \;.
\end{equation}
In electroweak precision analyses,
an error of $\Delta \mw\break
=40\mev$ is equivalent to an error
of $\Delta \mt=6\gev$, so increased precision for $\mw$ would
be of greatest immediate interest given the $\Delta\mt=4\gev$
error quoted above.
In order to make full use of the $\Delta\mt\lsim 0.5\gev$ precision 
possible at a $\mm$ collider would require $\Delta\mw\lsim 4\mev$.
We are currently studying the possibility that the latter can be
achieved at a $\mm$ collider.

Such precisions, combined with the essentially exact
determination of $\mh$ possible at a $\mm$ collider,
would allow a consistency test for precision electroweak measurements
at a hitherto unimagined level of accuracy.
If significant inconsistency is found, new physics could be revealed.
For example, inconsistency could arise if the light $\h$ is not
that of the SM but rather the $\hl$ of the MSSM and there is
a contribution to precision electroweak quantities
arising from the $\hh$ of the MSSM having a non-negligible $WW,ZZ$ coupling.
The contributions of stop and chargino states to loops would be another
example.

A precise determination of the top quark mass $\mt$ could well
be important in its own right.
One scenario is that the low-energy spectrum of particles
(SUSY or not) has been measured and there is a desert up to the GUT
scale. We would then want to extrapolate the low-energy parameters up to the
grand unified scale to test in a detailed way the physics at that scale. Then
the top quark mass (and the Yukawa coupling) would be crucially important
since this parameter determines to a large extent the evolution of all the
other Yukawas, including flavor mixings. These considerations become especially
important
if the top quark Yukawa coupling is determined by an infrared quasi-fixed point
for which very small changes in the top quark mass translate into very large
changes in the renormalized values of many other parameters in the theory.

\section{\uppercase{CP violation and FCNC in the Higgs Sector}}

A nonstandard Higgs sector could have sizable CP-violating effects as well
as new flavor changing neutral current (FCNC) effects that could be
probed with a $\mm$ collider. A general two Higgs doublet model has been
studied in Refs.~\cite{gg,as,pilaf}. There one would either
(i) measure correlations in
the final state, or (ii) transversely polarize the muon beams to observe
an asymmetry in the production rate as a function of spin orientation.
For the second option, the ability to achieve transverse polarization
with the necessary luminosity is a crucial consideration.

New FCNC effects could be studied as well\cite{ars}.
For example a Higgs in the
$s$-channel could exhibit the decay $\mm \to \hh\to t\overline{c}$.
This decay would have to compete against the $W\wstar$ decays.

\section{\uppercase{Exotic Higgs Bosons/Scalars}}

In general, a muon collider can probe any type of scalar that
has significant fermionic couplings.  Interesting new physics
could be revealed. To give one example, consider
the possibility that a doubly-charged Higgs boson with
lepton-number-violating coupling $\hmm\to \ell^-\ell^-$ exists,
as required in left-right symmetric models where the neutrino mass
is generated by the see-saw mechanism through a vacuum
expectation value of a neutral Higgs triplet field. 
Such a $\hmm$ could be produced in $\ell^-\ell^-$ collisions.
This scenario was studied in Ref.~\cite{ghmm} for an $e^-e^-$
collider, but a $\mu^-\mu^-$ collider would be even
better due to the much finer energy resolution (which enhances
cross sections) and the fact that the $\hmm\to\mu^-\mu^-$ coupling
should be larger than the $\hmm\to e^-e^-$ coupling.
%Further, a high luminosity $\mu^-\mu^-$ collider might be more
%easily developed as part of a muon collider facility than
%could an $e^-e^-$ collider at an electron collider facility.

Most likely, a $\hmm$ in the $\lsim500\gev$ region would already
be observed at the LHC by the time the muon collider
begins operation. In some scenarios, it would even be observed
to decay to $\mu^-\mu^-$ so that the required $s$-channel coupling would
be known to be non-zero.  However, the magnitude of the coupling would
not be determined; for this we would need the $\mu^-\mu^-$ collider.
In the likely limit where $\gamhmm\ll\sigrts$, the number of
$\hmm$ events for $L=50\fbi$ is given by
\begin{equation}
N(\hmm)=6\times 10^{11}\left({c_{\mu\mu}\over 10^{-5}}\right)
\left({0.01\%\over R(\%)}\right)\;,
\label{hmmrate}
\end{equation}
where the standard Majorana-like coupling-squared is parameterized as
\begin{equation}
|h_{\mu\mu}|^2=c_{\mu\mu} \mhmm^2(\gev)\;.
\end{equation}
Current limits on the coupling correspond to $c_{\mu\mu}\lsim 5\times 10^{-5}$.
Assuming that 30 to 300 events would provide a distinct
signal (the larger number probably required
if the dominant $\hmm$ decay channel is into $\mu^-\mu^-$, for
which there is a significant $\mu^-\mu^-\to \mu^-\mu^-$ background),
the muon collider would probe 
some 11 to 10 orders of magnitude more deeply in the 
coupling-squared than presently possible.  
This is a level of sensitivity
that would almost certainly be adequate for observing a $\hmm$ that is
associated with the triplet Higgs boson fields
that give rise to see-saw neutrino mass generation in the left-right symmetric
models.

\section{\uppercase{Physics at a 2{\protect\boldmath$\otimes$}2 TeV\hfill\break
{\protect\boldmath$\mu^+\mu^-$} Collider}}

Bremsstrahlung radiation scales like $m^{-4}$, 
so a circular storage ring can be
used for muons at high energies.
A high energy lepton collider with center-of-mass energy of $4\tev$ would
provide new physics reach beyond that contemplated at the LHC
or NLC (with $\rts\lsim 1.5\tev$). We concentrate primarily
on the following scenarios for physics at these energies: (1)~heavy
supersymmetric (SUSY) particles, (2)~strong scattering of longitudinal gauge
bosons (generically denoted\break $W_L$)
in the electroweak symmetry breaking\break (EWSB) sector, and (3)~heavy
vector resonance production, like a $Z'$.

\subsection{SUSY Factory}

Low-energy supersymmetry is a theoretically attractive extension of the 
Standard Model. Not only does it solve the
naturalness problem, but also the physics remains essentially perturbative
up to the grand unification scale, 
and gravity can be included by making the supersymmetry local.
Since the SUSY-breaking
scale and, hence, sparticle masses are required by naturalness
to be no larger than $1-2\tev$, a high energy $\mm$ collider 
with $\sqrt{s}=4\tev$ is guaranteed to be a SUSY factory if SUSY
is nature's choice.  Indeed, it may be the only machine that would
guarantee our ability to study the full spectrum of SUSY particles.
The LHC has sufficient energy to produce
supersymmetric particles but disentangling the spectrum and measuring the
masses will be a challenge  due to the complex cascade decays and QCD
backgrounds. The NLC would be a cleaner environment 
than the LHC to study the supersymmetric
particle decays, but the problem here may be insufficient energy to
completely explore the full particle spectrum.

\begin{figure}[h]
\let\normalsize=\captsize   %%%% changes the font to "\small"
\begin{center}
\centerline{\psfig{file=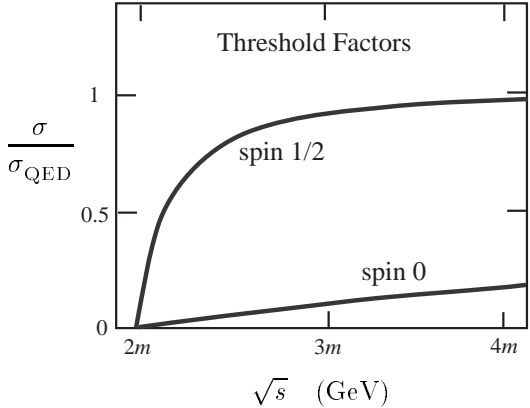,width=7cm}}
\begin{minipage}{7cm}       %%%% reduces width of caption to 7cm
%\smallskip
\caption{{\baselineskip=0pt
Comparison of kinematic suppression for
fermion pairs and squark pair production at $\ee$ or $\mm$ colliders.}}
\label{pwave}
\end{minipage}
\end{center}
\vspace{-.2in}
\end{figure}

Most supersymmetric models have a symmetry known as an $R$-parity that
requires that supersymmetric particles be created or destroyed in pairs.
This means that 
the energy required to find and study heavy scalars is more than twice
their mass. (If $R$-parity is violated, then sparticles can also
be produced singly; the single sparticle production rate would depend
on the magnitude of the violation, which is model- and generation-dependent.)
Further,  a $p$-wave suppression is 
operative for the production of scalars (in this case
the superpartners to the ordinary quarks and leptons), and energies well
above the kinematic threshold might be required to produce the scalar
pairs at an observable rate, as illustrated in Fig.~\ref{pwave}.
In addition, a large lever arm for exploring
the different threshold behaviour of spin-0 and spin-1/2 SUSY sparticles
could prove useful in mass determinations.

%10

To be more specific, it is useful to constrain the parameter
space by employing a supergravity (SUGRA) model. 
Such models are particularly attractive in that the breaking of the
electroweak symmetry is accomplished radiatively by the large top quark
Yukawa coupling driving one of the Higgs doublet masses negative through
renormalization group evolution. The simplest
SUGRA models contain the following parameters:
\begin{itemize}

\item a universal scalar mass $m_0$;

\item a universal gaugino mass $m_{1/2}$;

\item the ratio of the electroweak scale Higgs vev's, $\tan\beta=v_2/v_1$;

\item a universal trilinear term $A_0$;

\item the sign of the Higgs mixing: sign($\mu $).

\end{itemize}

The parameters above are constrained by various means.
Experimental bounds on the superpartner masses put a lower bound on $m_{1/2}$.
Naturalness considerations yield upper bounds on both $m_{1/2}$ and 
$m_0$, which, in turn, imply upper limits on the 
superparticle masses. If one supposes that the LSP is 
the cold dark matter of the universe, then there is an upper limit on $m_0$
so that the annihilation channels for the LSP are not suppressed by the 
heavy scalar masses. The $A_0$ parameter is limited by the requirement
of an acceptable vacuum state; $1\lsim\tanb\lsim 50-60$ is required
for perturbativity of the Yukawa couplings. A representative
choice of parameters
that is consistent with all these constraints, but at the same time
illustrates the power of a $\mm$ collider is:
\begin{eqnarray}
& m_0=2m_{1/2}=500\gev\;,  \nonumber \\
& \tan \beta=2,\quad A_0=0,\quad \mu <0 \;.
\label{sugraparameters}
\end{eqnarray}
By adopting a large ratio of $m_0/m_{1/2}=2$
the scalars become heavy (with the exception of the
lightest Higgs boson) compared to the gauginos.
The particle and sparticle masses obtained from
renormalization group evolution are:
\begin{eqnarray}
& \mhl=88\gev,\quad \mha=921\gev\;, \\  
& m_{H^\pm}=\mhh=924\gev\;, \\
& m_{\wtil q_L}\simeq 752\gev,\quad m_{\wtil q_R}\simeq 735\gev\;,\\
& m_{\wtil b_1}=643\gev,\quad m_{\wtil b_2}=735\gev\;,\\
& m_{\wtil t_1}=510\gev,\quad m_{\wtil t_2}=666\gev\;,\\
& m_{\wtil \nu}\sim m_{\wtil\ell}\sim 510-530\gev\;,\\
& m_{\wtil \chi^0_{1,2,3,4}}=107,217,605,613\gev\;,\\
& m_{\wtil \chi^+_{1,2}}=217,612\gev\;.
\label{sugramasses}
\end{eqnarray}
Thus, the choice of GUT parameters, Eq.~(\ref{sugraparameters}), 
leads, as desired, to a scenario such that
pair production of heavy scalars is only accessible at a high
energy machine like the NMC.

\begin{figure}
\let\normalsize=\captsize   %%%% changes the font to "\small"
\begin{center}
%\centerline{\psfig{file=eehh.ps,width=7cm}}
\centerline{\psfig{file=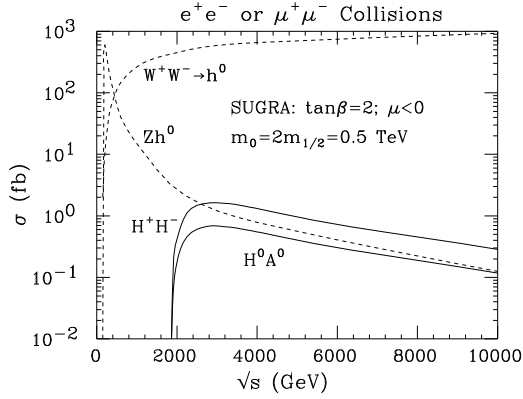,width=7.5cm}}
\begin{minipage}{7cm}       %%%% reduces width of caption to 7cm
\smallskip
\caption{{\baselineskip=0pt
Pair production of heavy Higgs bosons at a high energy lepton collider.
For comparison, cross sections for the lightest Higgs boson production 
via the Bjorken process $\mm\to \zstar\to Z\hl$ and
via the $WW$ fusion are also presented.}}
\label{eehh}
\end{minipage}
\end{center}
\vspace{-.1in}
\end{figure}

First, we consider the pair production of the heavy Higgs bosons
\begin{eqnarray}
\mm &\to & Z \to \hh\ha\;, \\
\mm &\to & \gamma,Z \to H^+H^-\;.
\end{eqnarray}
The cross sections are shown in Fig.~\ref{eehh} versus $\rts$. 
A $\mm$ collider with $\rts \gsim 2\tev$ is needed and 
well above the threshold the cross section is ${\cal O}(1\fb)$.
In the scenario of Eq.~(\ref{sugraparameters}), the decays of these
heavy Higgs bosons are predominantly into top quark
modes ($t\overline{t}$ for the
neutral Higgs and $t\overline{b}$ for the charged Higgs), with branching
fractions near 90\%. Observation of the $\hh$, $\ha$, and $\hpm$
would be straightforward even for a pessimistic luminosity of $L=100\fbi$.
Backgrounds would be negligible once the requirement of roughly equal masses
for two back-to-back particles is imposed.

In other scenarios the decays may be more complex and
include multiple decay modes into supersymmetric particles,
in which case the overall event rate might prove crucial to establishing a
signal. In some scenarios investigated in Ref.~\cite{gk} complex decays
are important, but the $\mm$ collider has sufficient production
rate that one or more of the modes
\begin{eqnarray}
&&(\hh\to b\overline{b})+(\ha\to b\overline{b})\;, \\
&&(\hh\to \hl\hl\to b\overline{b}b\overline{b})+(\ha\to X)\;, \\
&&(\hh\to t\overline{t})+(\ha\to t\overline{t})\;,
\end{eqnarray}
are still visible above the backgrounds for $L\gsim 500\fbi$.
Despite the significant dilution of the
signal by the additional SUSY decay modes (which is most important at low
$\tan \beta$), one can observe a signal of $\gsim 50$~events in one channel or
another.

%11

The high energy $\mm$ collider will yield a large number of 
the light SM-like $\hl$ via $\mm\to \zstar\to Z\hl$ and
$WW$ fusion, $\mm\to \nu\overline{\nu}\hl$.
In contrast to a machine running at FMC
energies ($\rts\sim 500\gev$), where the cross sections for these
two processes are comparable, at higher
energies, $\rts\gsim 1\tev$, the $WW$
fusion process dominates as shown in Fig.~\ref{eehh}.

Any assessment of the physics signals in the pair production of the 
supersymmetric partners of the quarks and leptons is model-dependent.
However, as illustrated by the specific SUGRA scenario masses
of Eq.~(\ref{sugramasses}), squarks are expected to be 
somewhat heavier than the sleptons due to their QCD interactions which affect
the running of their associated `soft' masses
away from the universal mass $m_0$ in the evolution 
from the GUT scale to low energies. Except for the LSP,
the lightest superpartner of each type decays to a gaugino (or gluino) and 
an ordinary fermion, and the gaugino will decay if it is not the LSP.
Since the particles are generally 
too short-lived to be observed, we must infer everything about their 
production from their decay products. 

\begin{figure}[h]
\let\normalsize=\captsize   %%%% changes the font to "\small"
\begin{center}
%\centerline{\psfig{file=susyprod.ps,width=cm}}
\centerline{\psfig{file=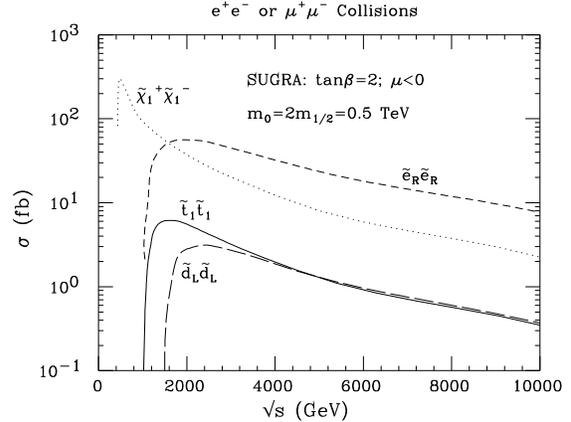,width=7.5cm}}
\begin{minipage}{7cm}       %%%% reduces width of caption to 7cm
\smallskip
\caption{{\baselineskip=0pt
The production cross sections for SUSY particles
in a supergravity model with heavy scalars.}}
\label{susyprod}
\end{minipage}
\end{center}
\end{figure}

We illustrate
the production cross sections for several important sparticle pairs in 
Fig.~\ref{susyprod} for the SUGRA model of Eq.~(\ref{sugraparameters}).
For a collider with $\rts \sim 4\tev$, cross
sections of $\sim 2$--30~fb are expected.

%12

The final states of interest are determined by the dominant
decay modes, which in this model
are $\wtil e_R\to e\wtil \chi_1^0$ ($BF=0.999$), 
$\wtil\chi_1^+\to \wp \wtil\chi_1^0$ ($BF=0.999$), 
$\wtil d_L\to \wtil \chi_1^- u,\wtil\chi_2^0 d,\wtil g d$
($BF=0.52,0.27,0.20$), and $\wtil t_1\to \wtil\chi_1^+ t$.  Thus, for example,
with a luminosity of $L=200\fbi$ at $\rts=4\tev$, 
$\wtil d_L$ pair production would result in $200\times 2\times (0.52)^2=
100$ events containing two $u$-quark jets, two energetic leptons
(not necessarily of the same type), and substantial missing energy.
The SM background should be small, and the signal would be clearly visible.
The energy spectra of the quark jets would allow a determination
of $m_{\wtil d_L}-m_{\wtil\chi_1^+}$ while the lepton
energy spectra would fix $m_{\wtil \chi_1^+}-m_{\wtil\chi_1^0}$.
If the machine energy can be varied, then the turn-on of such
events would fix the $\wtil d_L$ mass.  The $\wtil\chi_1^+$
and $\wtil\chi_1^0$ masses would presumably already be known from
studying the $\ell^+\ell^-+$missing-energy
signal from $\wtil\chi_1^+\wtil\chi_1^-$ pair production, best
performed at much lower energies.
Thus, cross checks on the gaugino masses are possible, while
at the same time two determinations of the $\wtil d_L$ mass
become available (one from threshold location
and the other via the quark jet spectra combined
with a known mass for the $\wtil\chi_1^+$). 

This example illustrates the power of a $\mm$
collider, especially one whose energy can be varied over a broad range.
Maintaining high luminosity over a broad energy range may require the
construction of several (relatively inexpensive) final storage rings.

%\subsection{$\boldmath W_LW_L\to W_LW_L$ to probe EWSB}
\subsection{The {\protect\boldmath $W_LW_L\to W_LW_L$} probe of EWSB}

A compelling motivation for building any new machine is to discover the
mechanism behind\break
EWSB. This may involve directly producing the Higgs particle
of the Standard Model or supersymmetric particles.
Alternatively it could be that no light Higgs bosons exist;
then general arguments based on partial wave unitarity require that the
interactions of the longitudinal gauge bosons ($W$ and $Z$) become
strong and nonperturbative. The energy
scale where this happens is about 1--2~TeV,
implying that a collider needs to probe vector boson scattering
at energies at least this high. The LHC energy and 
the currently envisioned NLC energies (up to $\sim 1.5\tev$) are 
marginally able to do this.
In contrast, a $4\tev$ muon collider is in the
optimal energy range for a study of strong vector boson scattering.
The construction of a multi-TeV $\ee$ collider is also a 
possibility\cite{dburke}.)

%13
\begin{figure}
\let\normalsize=\captsize   %%%% changes the font to "\small"
\begin{center}
\centerline{\psfig{file=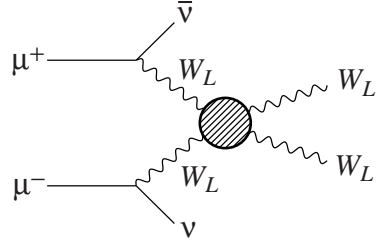,width=5cm}}
\begin{minipage}{7cm}       %%%% reduces width of caption to 7cm
\bigskip
\caption{{\baselineskip=0pt
Symbolic diagram for strong WW scattering.}}
\label{wwscatfig}
\end{minipage}
\end{center}
\end{figure}

Strong electroweak scattering (SEWS) effects 
can be estimated by using 
the Standard Model with a heavy Higgs as a prototype of
the strong scattering sector. The SM with a light Higgs 
is an appropriate definition of the electroweak background
since only transversely polarized $W$'s contribute
to vector boson scattering when the Higgs has a small mass. 
For a 1 TeV SM Higgs boson, the signal is thus defined as
\begin{equation}
\Delta \sigma =\sigma(\mhsm=1\;{\rm TeV})-\sigma(\mhsm=10\;{\rm GeV})\;.
\end{equation}
Results for $\Delta\sigma$ are shown in Table~\ref{tableii}
for $\rts=1.5\tev$ (possibly the upper limit
for a first $\ee$ collider) and $4\tev$. The strong 
scattering signal is relatively small at energies of order $1\tev$, but 
grows substantially as multi-TeV energies are reached.
Thus, the highest energies in $\sqrt{s}$ that can
be reached at a muon collider could be critically important.

\begin{table}[h]
\centering
\caption[]{\label{tableii}\small
Strong electroweak scattering signals in $\wp\wm\to\wp\wm$
and $\wp\wm\to ZZ$ at future lepton colliders.}
\medskip
\begin{tabular}{|c|c|c|}
\hline
$\sqrt s$& $\Delta\sigma(W^+W^-)$& $\Delta\sigma(ZZ)$\\ \hline \hline
1.5 TeV& 8 fb& 6 fb\\ \hline
4 TeV& 80 fb& 50 fb
\\ \hline
\end{tabular}
\end{table}

Many other models   for the strongly interacting gauge
sector have been constructed in addition to the SM, including\cite{bbcghlry}:
\begin{itemize}

\item a (``Scalar'') model in
which there is a scalar Higgs resonance with $M_S=1\tev$ but non-SM width
of $\Gamma_S=350\gev$;
\item a (``Vector'') model in which there is no scalar resonance,
but rather a vector resonance with $M_V=1\tev$ and $\Gamma_V=35\gev$;
\item a model, denoted by ``LET'' or ``$\mhsm=\infty$'', in which 
the SM Higgs is taken to have infinite mass and the partial waves simply
follow the behavior predicted by the low-energy theorems;
\item a model (denoted by ``LET-K'') in which the LET behavior is unitarized
via $K$-matrix techniques.
\end{itemize}
To differentiate among models,
a complete study of the physics of strongly interacting gauge bosons would
be required. In particular, all the following vector-boson
scattering channels must be studied:
\begin{eqnarray}
W^+W^-&\to &W^+W^-,ZZ\;, \\
W^\pm Z&\to &W^\pm Z\;, \\
W^\pm W^\pm &\to &W^\pm W^\pm \;.
\end{eqnarray}
Partial exploration of the three isospin channels can be made at the LHC.
The signal and background for gold-plated (purely leptonic) events is shown
in Table~\ref{tableiii} for the LHC operating at $14\tev$ with $L=100\fbi$,
for several of the above models.
These channels have also been studied for a $1.5\tev$ NLC\cite{bchp},
and, again, event rates are at a level that
first signals of the strongly interacting vector boson
sector would emerge, but the ability to discriminate between models
and actually study these strong interactions would be limited.
\begin{table}[h]
\centering
\caption[]{\label{tableiii}\small
Total numbers of $W_LW_L \to 4$-lepton
signal $S$ and background $B$ events calculated for the
LHC\protect\cite{bbcghlry}, assuming $L=100\fbi$.}
\medskip
\begin{tabular}{|c|c|c|c|c|}
\hline
& Bkgd & Scalar & Vector & LET-K \\ 
\hline \hline
$ZZ (4\ell)$& 1 & 5 & 1.5 & 1.5\\
$(2\ell 2\nu)$& 2 & 17 & 5 & 4.5\\ \hline
$W^+W^-$& 12 & 18 & 6 & 5\\ \hline
$W^+Z$& 22 & 2 & 70 & 3\\ \hline
$W^{\pm }W^{\pm }$& 4 & 7 & 12 & 13\\ \hline
\end{tabular}
\end{table}

For a $\mm$ collider operating at $4\tev$ the statistical significances
markedly improve. Table~\ref{tableiv} summarizes the total signal $S$ and
background $B$
event numbers, summing over diboson invariant mass bins,  together with
the statistical significance $S/\sqrt B$ for different models of the 
strongly-interacting physics. 
A broad Higgs-like scalar will enhance both $W^+ W^-$
and $ZZ$ channels with $\sigma(W^+ W^-) > \sigma(ZZ)$; a $\rho$-like vector
resonance will manifest itself through $W^+W^-$ but not $ZZ$; while the 
$\mhsm=\infty$ (LET)
amplitude will enhance $ZZ$ more than $W^+ W^-$.
The  $\mhsm=\infty$
signal for  $W^+W^-$ is visible, although still far from robust;
the ratio $S/B$ can be enhanced by making a higher
mass cut (\eg\ $M_{WW} > 0.7$ TeV), but the significance $S/\sqrt B$
is not improved.

%14
\begin{figure}[h]
\let\normalsize=\captsize   %%%% changes the font to "\small"
\centering
\centerline{\psfig{file=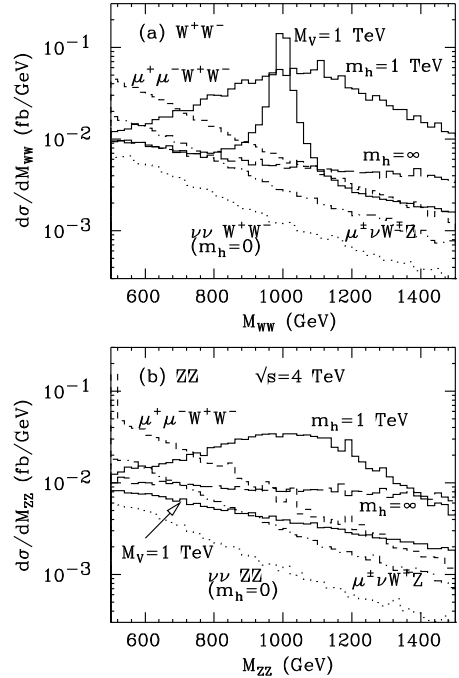,width=6cm}}
\begin{minipage}{7cm}       %%%% reduces width of caption to 7cm
\smallskip
\caption{{\baselineskip=0pt
Histograms for the signals and backgrounds in strong 
vector boson scattering in the 
(a) $\wp\wm$ and (b) $ZZ$ final states. The background is given by the 
strictly electroweak $\mhsm=0$ limit  of the Standard Model. The 
three signals shown
are (I) a vector resonance with $M_V=1\tev$, $\Gamma_V =35\gev$, (II) 
the SM Higgs with $\mhsm=1\tev$, and (III) the SM with $\mhsm=\infty$
(LET model). In the figure the shorthand notation $h$ is used for $\hsm$.}}
\label{wwhistograms}
\end{minipage}
\end{figure}

\begin{table*}
\centering
\caption[]{\label{tableiv}\small
Total numbers of $W^+W^-, ZZ \rightarrow  4$-jet
signal $S$ and background $B$ events calculated for  a 4~TeV
$\protect \mm$ collider
with  integrated luminosity 200~fb$^{-1}$.  Events are summed
over the mass range $0.5 < M_{WW} < 1.5$~TeV except for the $W^+W^-$ channel
with  a narrow vector resonance for which $0.9 < M_{WW} < 1.1$~TeV. The
statistical significance $S/\sqrt B$ is also given.
The hadronic branching fractions of $WW$ decays and the $W^\pm/Z$
identification/misidentification are included.}
\medskip
\begin{tabular}{|l|c|c|c|c|c|} \hline
channels & SM  & Scalar & Vector   & SM  \\
\noalign{\vskip-1ex}
& $\mhsm=1$ TeV & $M_S=1$ TeV & $M_V=1$ TeV & $\mhsm=\infty$ \\
\hline
$S(\mu^+ \mu^- \to \bar \nu \nu W^+ W^-)$
& 1900   & 1400   & 370  & 230  \\
$B$(backgrounds)
& 1100    & 1100   & 110  & 1100  \\
$S/\sqrt B$ & 57 & 42 & 35 & 6.9 \\
\hline
$S(\mu^+ \mu^- \to \bar\nu \nu ZZ)$
&  970  & 700  & 220  & 350   \\
$B$(backgrounds)
& 160    & 160   & 160  & 160  \\
$S/\sqrt B$ & 77& 55& 17& 28\\
\hline
\end{tabular}
\end{table*}

Signals and the irreducible electroweak background for the $\wp\wm$
and $ZZ$ modes are
shown in Fig.~\ref{wwhistograms}. The complementarity of these two modes is 
clear from the figure. However,
to make use of this complementarity it is crucial
to be able to distinguish final state 
$W$ and $Z$ bosons using the dijet invariant masses. This is possible 
provided there is sufficient jet energy resolution, as discussed in 
Ref.~\cite{bchp}.

Finally, we note that event numbers in the 1~TeV SM Higgs 
and Vector resonance cases, and possibly
even in the $\mhsm=\infty$ (LET) case,
are such that not only could a substantial
overall signal be observed, but also at high $L$ 
the shape of the excess, due to strong interactions,
in the distribution in vector boson pair mass 
could be measured over a broad interval in the $1\tev$ range.  
For instance, from Fig.~\ref{wwhistograms}a
in the case of $\mhsm=\infty$,
a 100 GeV interval from 1.4 TeV to 1.5 TeV would contain
$L\times 100\gev\times (4\times 10^{-3}\fb/\gev)=400$
signal events for $L=1000\fbi$, thereby allowing a 5\% measurement
of the $m_{\wp\wm}$ signal distribution in this bin. 
The level of accuracy in this one bin alone 
would distinguish this model from the Vector or $\mhsm=1\tev$ models.
The difference between the three different distributions
plotted in Fig.~\ref{wwhistograms} could be tracked in both channels.
The ability to measure the distributions with reasonable precision 
would allow detailed insight into
the dynamics of the strongly interacting electroweak sector
when the collider achieves energies substantially
above $1\tev$. Thus, if some signals for a strongly interacting
sector emerge at the LHC, a $\rts = 3-4\tev$ $\mm$ (or $\ee$, if possible)
collider will be essential.

\subsection{Exotic Heavy States}

The very high energy of a $4\tev$ collider would open up the possibility
of directly producing many new particles outside of the Standard Model.
Some exotic heavy particles that could be discovered and studied at a muon
collider are (1) sequential fermions, $Q\overline{Q}$, 
$L\overline{L}$\cite{gmp},
(2) lepto-quarks, (3)~vector-like
fermions\cite{bbp}, and (4) new gauge bosons like a $Z'$ or $W_R$\cite{hrprep}.

%15
\begin{figure}[h]
\let\normalsize=\captsize   %%%% changes the font to "\small"
\begin{center}
\centerline{\psfig{file=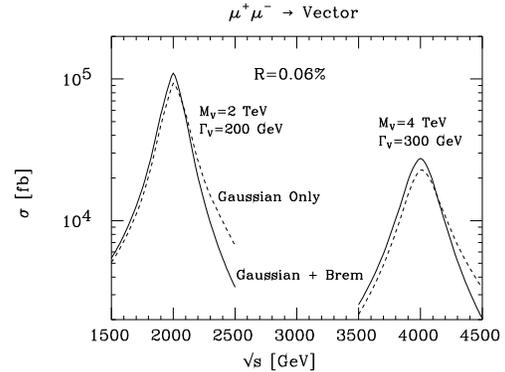,width=7cm}}
\begin{minipage}{7cm}       %%%% reduces width of caption to 7cm
\bigskip
\caption{{\baselineskip=0pt
High event rates are possible if the muon collider energy is set equal to the
vector resonance ($Z'$ or $\rho _{\rm TC}$) mass. Two examples are shown here
with $R=0.06\%$.}}
\label{vectonres}
\end{minipage}
\end{center}
\end{figure}

\begin{figure}[t]
\let\normalsize=\captsize   %%%% changes the font to "\small"
\begin{center}
%\centerline{\psfig{file=rtstail_spectrum.ps,width=7cm}}
\centerline{\psfig{file=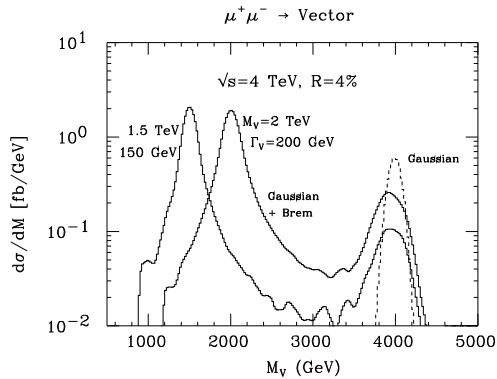,width=7cm}}
\begin{minipage}{7cm}       %%%% reduces width of caption to 7cm
\bigskip
\caption{{\baselineskip=0pt
A heavy vector resonance can be visible in the bremsstrahlung tail of a 
high energy collider. Here a $\mm$ collider operating at $4\tev$ is shown 
for $M_V=1.5\tev$ and $2\tev$.}}
\label{vectoffres}
\end{minipage}
\end{center}
\end{figure}

A new vector resonance such as a $Z'$ or a technirho, $\rho _{\rm TC}$, is a 
particularly interesting
possibility. The collider could be designed to sit on the
resonance $\rts \sim M_V$ in which case it would function as a $Z'$ or
$\rho _{\rm TC}$ factory as illustrated in Fig.~\ref{vectonres}.
Alternatively, if the mass of the resonance is not
known a priori, then the collider operating at an energy above 
the resonance mass could discover it via
the bremsstrahlung tail shown in Fig.~\ref{bremtail}.
Figure~\ref{vectoffres}
shows the differential cross section in the reconstructed
final state mass $M_V$
for a muon collider operating at $4\tev$ for two cases where the vector
resonance has mass $1.5\tev$ and $2\tev$.
Dramatic and unmistakable signals would appear
even for integrated luminosity as low as $L\gsim 50-100\fbi$.

\section{\uppercase{Conclusions}}

%16

A muon collider is very likely to 
add substantially to our knowledge of physics in the
coming decades. A machine with energy in the range $\rts=100$--500~GeV is
comparable to the NLC and provides valuable
additional features. The most notable of
these is the possibility of creating a Higgs boson in the
$s$-channel and
measuring its mass and decay widths directly and precisely.
Even if a light Higgs does not exist,
studies of the $t\anti t$ and $\wp\wm$ thresholds at such a low-energy machine
would yield higher precision in determining $\mt$ and $\mw$ than possible
at other colliders.
A $\mm$ collider with energy as high as $\rts \sim 4\tev$ appears
to be entirely feasible and is ideally suited for studying a 
strongly-interacting symmetry breaking sector, 
since the center-of-mass energy is well
above the energy range at which vector
boson interactions must become strong. Many other 
types of exotic physics beyond the
Standard Model could be probed at such a high machine energy.
For example, if supersymmetry exists, a $4\tev$ $\mm$ collider would be a factory
for sparticle pair production.  Observation of a heavy $Z^\prime$ 
in the bremsstrahlung luminosity tail
would be straightforward and the machine energy
could later be reset to provide a $Z^\prime$ factory.
All the issues presented in this paper will be discussed in greater
detail in a forthcoming review article\cite{mupmumreport}.

\section*{\uppercase{Acknowledgments}}

This work was supported in part by the U.S.\break
Department of Energy  
under Grants No.\ DE-\break
FG02-95ER40896, No.~DE-FG03-91ER40674 and  
No.~DE-FG02-91ER40661. 
Further support was provided
by the University of Wisconsin Research
Committee, with funds granted by the Wisconsin Alumni Research  
Foundation, and by the Davis Institute for High Energy Physics.

\end{document}